\documentclass[10pt]{article}
\usepackage{fullpage,graphicx,subfigure,mathdots,mathpazo,color}
\usepackage{amsmath,amscd,tikz,mathrsfs,cite}
\usepackage[normalem]{ulem}
\usepackage{amsmath}
\usepackage{setspace}

\usepackage{epsfig,amsmath,graphicx,amssymb,overpic}
\usepackage{bm}
\usepackage{graphicx}
\usepackage{subfigure, xcolor}
\usepackage{ntheorem}
\usepackage{listings,diagbox}

\usepackage{color}
\usepackage{epsfig,amsmath,graphicx,amssymb,overpic}

\usepackage{booktabs}
\usepackage{diagbox}
\usepackage{graphicx}
\usepackage{dcolumn}
\usepackage{bm}
\usepackage{graphicx}
\usepackage{subfigure}
\usepackage{epsfig,amsmath,graphicx,amssymb,overpic,cite}
\usepackage{makecell}

\def\be{\begin{equation}}
\def\ee{\end{equation}}
\def\bee{\begin{eqnarray}}
\def\ene{\end{eqnarray}}
\def\bes{\begin{subequations}}
\def\ees{\end{subequations}}

\newcommand{\PT}{{\cal PT}}

\def\v{\vspace{0.1in}}

\usepackage{geometry}
\allowdisplaybreaks[4]

\begin{document}

\baselineskip=14pt
\renewcommand {\thefootnote}{\dag}
\renewcommand {\thefootnote}{\ddag}
\renewcommand {\thefootnote}{ }

\pagestyle{plain}

\begin{center}
\baselineskip=16pt \leftline{} \vspace{-.3in} {\Large \bf Data-driven soliton mappings for
integrable fractional nonlinear \\ wave equations via deep learning with Fourier neural operator} \\[0.2in]
\end{center}

\begin{center}
Ming Zhong$^{a,b}$ and Zhenya Yan$^{{a,b},{*}}$ \footnote{$^{*}$Corresponding author. {\it Email address}: zyyan@mmrc.iss.ac.cn}  \\[0.15in]
{\it \small$^{a}$KLMM, Academy of Mathematics and Systems Science, Chinese Academy of Sciences, Beijing 100190, China \\
\it \small$^{b}$School of Mathematical Sciences, University of Chinese Academy of Sciences, Beijing 100049, China} \\
\end{center}

\vspace{0.1in}

{\baselineskip=13pt


\vspace{0.08in}


\noindent {\bf Abstract:}\, In this paper, we firstly extend the Fourier neural operator (FNO) to discovery the soliton mapping
between two function spaces, where one is the fractional-order index space $\{\epsilon|\epsilon\in (0, 1)\}$ in the fractional integrable nonlinear wave equations while another denotes the solitonic solution function space. To be specific, the fractional nonlinear Schr\"{o}dinger (fNLS), fractional Korteweg-de Vries (fKdV), fractional modified Korteweg-de Vries (fmKdV) and fractional sine-Gordon (fsineG) equations proposed recently are studied in this paper. We present the train and evaluate progress by recording the train and test loss. To illustrate the accuracies, the data-driven solitons are also compared to the exact solutions. Moreover, we consider the  influences of several critical factors (e.g., activation functions containing Relu$(x)$, Sigmoid$(x)$, Swish$(x)$ and $x\tanh(x)$, depths of fully connected layer) on the performance of the  FNO algorithm. We also use a new activation function, namely, $x\tanh(x)$, which is not used in the field of deep learning. The results obtained in this paper may be useful to further understand the neural networks in the fractional integrable nonlinear wave systems and the mappings between two spaces.

\vspace{0.1in} \noindent  {\it Keywords:} \,\,  Fractional integrable nonlinear equations; \,
Fourier neural operator,\, deep learning;\,  data-driven soliton mapping;\, activation function;\, depth of fully connected layer


\vspace{-0.05in}

\vspace{0.1in}

\section{Introduction}

\quad Integrable nonlinear systems and soliton theory  are of important significance in the field of nonlinear science~\cite{solitonbook1,solitonbook2,solitonbook3,solitonbook4,solitonbook5,solitonbook6}. The discovery of solitons can be traced back as far as the shallow-water solitary waves recorded by Russell in 1834~\cite{KDV1}, and then described by
the well-known KdV equation established by Korteweg and de Vries in 1895~\cite{KDV2}. It has drawn the attention
of numerous researchers due to its applications in many fields, such as classical mechanics, fluid mechanics, elastic mechanics, quantum mechanics, canonical field theory, optics, and biology~\cite{apply1,apply2,apply3,apply4,apply5,apply6,apply7}. Due to its  significance in the nonlinear area, various effective analytical methods have been established and developed for such nonlinear soliton equations, such as Hirota bilinear method~\cite{Hirota1}, Darboux transform~\cite{Darboux1,Darboux2}, inverse scattering transform (IST)~\cite{Inverse1,Inverse2}, and so on~\cite{RH,Tanh}.  And since the stably propagating optical solitons were proved to appear in $\PT$-symmetric dissipative systems~\cite{pt1}, the theory of solitons in $\PT$ nearly-integrable systems has also been studied extensively~\cite{pt2,pt3,pt8}.  The $\PT$-symmetric structure also generalizes the concept of solitons, which were previously thought to exist only in conservative systems.

Recently, the fractional integrable system was firstly put forward by Ablowitz and his collaborators~\cite{fraction1}. They derived the fractional nonlinear Schr\"{o}dinger (fNLS) and fractional Korteweg-de Vries (fKdV) equations, and their corresponding one-soliton solutions by  utilizing the completeness relations, dispersion relations, and IST. The completeness relations can be used to verify the correctness of the fNLS and fKdV equation. In fact, compared with the usual NLS and KdV equations,
the  spatial  part of the so-called formal Lax pair
remains unchanged, while the  temporal part of the formal Lax pair cannot be obtained explicitly, but the soliton solution can be derived through the dispersion relation and inverse scattering  transform. It is worth noting that the fractional order derivatives mentioned in this paper are in the sense of the Riesz fractional derivative~\cite{Rises1,Rises2}. The fKdV equation can be written as the form of
\begin{equation}\label{fKdV}
u_{t}+\gamma(L^{A}) u_{x}=0, \quad L^{A} \equiv-\frac{1}{4} \partial_{x}^{2}-u+\frac{1}{2} u_{x} \int_{x}^{\infty} d y
\end{equation}
with $\gamma(L^{A})=-4 L^{A}\left|4 L^{A}\right|^{\epsilon}$ and $\epsilon\in(0,1)$ (similar hereinafter). In particular, as $\epsilon=0$, the fKdV equation reduces to the KdV equation
\bee
 u_t+6uu_x+u_{xxx}=0.
\ene
The subscript denotes the derivative with respect to time ($t$) or the spatial propagation coordinate ($x$). 
The fNLS can also derive as the second component  of
\begin{equation}\label{fNLS}
\sigma_{3} \partial_{t} \mathbf{u}+2 A_{0}(\mathbf{L}^{A}) \mathbf{u}=0, \quad \sigma_{3}=\left(\begin{array}{cc}
1 & 0 \\
0 & -1
\end{array}\right),
\end{equation}
where $\mathbf{u}=(\phi, \psi)^{T}$ and the operator
\begin{equation}\label{fNLSoperator}
\mathbf{L}^{A} \equiv \frac{1}{2 i}\left(\begin{array}{cc}
\partial_{x}-2 \phi I_{-} \psi & 2 \phi I_{-} \phi \v\\
-2 \psi I_{-} \psi & -\partial_{x}+2 \psi I_{-} \phi
\end{array}\right),
\end{equation}
with $I_{-}=\int_{-\infty}^{x} d y$ and $A_{0}\left(\mathbf{L}^{A}\right)=2 i\left(\mathbf{L}^{A}\right)^{2}\left|2 \mathbf{L}^{A}\right|^{\epsilon}$, and $\phi=\mp \psi^*$, where $*$ denotes the  complex conjugate. In particular, as $\epsilon=0$, the fNLS equation reduces to the NLS equation
\bee
 i\psi_t-\psi_{xx}\mp|\psi|^2\psi=0.
\ene

 After that,  Ablowitz {\it et al}~\cite{fraction2} also discussed the fmKdV and fsineG  equations:
\begin{equation}\label{fmkdv}
\begin{gathered}
\text{fmKdV}:\quad u_{t}-4 L_{\pm}^{A}\left|2 L_{\pm}^{A}\right|^{\epsilon} u_{x}=0, \\
\text{fsineG}:\quad  u_{t x}+\frac{\left|4 L_{-}^{A}|_{u\to -u_x/2}\right|^{\epsilon}}{4 L_{-}^{A}|_{u\to -u_x/2}} u_{x x}=0 \\
\end{gathered}
\end{equation}
with $L_{+}^{A}=-\frac{1}{4} \partial_{x}^{2}+u^{2}+u_{x} I_{-} u$ and $L_{-}^{A}=-\frac{1}{4} \partial_{x}^{2}-u^{2}-u_{x} I_{-}u$.

Recently, with the rapid development of  hardware (GPU) and the continuous updating of deep learning (DL)  libraries (e.g., Pytorch~\cite{torch}, Tensorflow~\cite{tf}, Julia~\cite{julia}), DL  becomes a hot spot for research gradually.  The trainable parameters $\Theta$ (weights $w$, bias $b$) in the neural network provides a large degree of freedom for the approximation capability of the network, thus DL can extract features from  large amounts of data. Nowadays, DL has been effectively deployed in a variety of fields, such as image recognition~\cite{Kr12,He16}, target detection~\cite{Gi15}, image generation~\cite{Ar17}, machine translation~\cite{Va17}, speech recognition~\cite{Hi12}, natural language processing~\cite{De19}, and etc.
Although deep learning has achieved remarkable achievements in such fields, only recently has it been gradually introduced into the field of scientific computing~\cite{sciml}. Unlike traditional network-based numerical methods, deep learning methods are gridless methods that can somewhat break the problems of  dimensional catastrophe for traditional formats. Due to the existence of the universal approximation theorem~\cite{ut1,ut2}, it is a natural notion to utilize neural networks to learn ordinary differential equations (ODEs) or partial differential equations (PDEs).

Deep learning-based ODEs/PDEs solvers can be broadly divided into two schools of thought. One of which is the utilization of linear/nonlinear combinations of neural networks as basis functions for the approximation functions,  such as the physics-informed  neural networks (PINNs) method~\cite{Ra19,pinn,PDE}, the deep Ritz method~\cite{e-2018}. It can be used to  exploit measurements on collocation points to approximate PDE solutions. The other is the DL-based neural operator which aims to learn the mapping between two infinite Banach space,  such as the DeepOnet~\cite{deeponet}, Fourier neural operator (FNO)~\cite{fno1,fno2}, and other neural operator methods~\cite{NO-2020,NO-2021}. The former focuses on the solving of the particular instance issue, even if there are minor changes in the equations or initial/boundary conditions that require re-training and prediction, while the latter concentrates on learning the mapping of two infinite-dimensional spaces, and can evaluate numerous instances without re-training. Although the latter is more demanding in terms of network generalization, it is more valuable for applications. The most distinctive feature of FNO is the introduction of the Fourier layer through the Fourier transform, which greatly enhanced the ability of the network to extract features~\cite{fno-th}. And for the given PDE families under the certain distribution, FNO can be quickly given for any given new parameters once the networks are trained. Given the powerful approximation capability of FNO, it has been applied to a number of fields, such as weather forecasting~\cite{Yi22}, the multi-phase flow~\cite{We22}, and heterogeneous material modeling~\cite{Yo22}, and soliton equations~\cite{zhong22}. Nowadays, many variants have been derived to further improve the generalization ability of FNO~\cite{fnov1,fnov2,fnov3}.

As mentioned above, FNO was first proposed for solving the forward problem of parametric PDE. In fact, FNO aims to learn the relations between two function spaces from the observation. Motivated by these existing results, we extend FNO to learn the mapping between the fractional-order index $\epsilon$ and the soliton function space for fractional integrable nonlinear wave equations. In other words, the FNO can give the soliton solutions in the whole spario-temporal area for the input $\epsilon$. The above framework allows us to move away from the traditional framework of the forward problem for PDEs to the study of the relationships between systems.

The rest of this paper is arranged as follows. In Sec. 2, we specifically give the neural network framework of the FNO method, and the choice of some hyper-parameters. In Sec. 3, we learn the mapping between the fractional-order index $\epsilon\in (0,1)$ to
the soliton function space for the above fractional integrable nonlinear systems such as fKdV, fNLS, fmKd and fsineG equations.
 In Sec. 4, several critical factors are considered to have the effect on the FNO scheme in the case of fKdV equation, such as the nonlinear activation functions and depths for the fully connected layer $P$. Finally, we give some conclusions and discussions in Sec. 5.

\section{Deep learning FNO methodology}
The deep learning FNO focuses on the learning a nonlinear mapping $\mathcal{G}$: $\mathcal{H}_1\rightarrow \mathcal{H}_2$ with an underlying bounded spacial domain $D\subset \mathcal{R}^2$. The space $\mathcal{H}_1$ and $\mathcal{H}_2$ denote the parametric space (fraction order) and soliton space respectively. And the domain should be  discretized  for the input and output observations  in the training and evaluate process. By introducing the neural network, the FNO parameterizes the nonlinear mapping $\mathcal{G}$ into $\mathcal{G}_\Theta$:
\begin{equation}\label{fno}
  \mathcal{G}_\Theta: \mathcal{H}_1\times \Theta\rightarrow \mathcal{H}_2,
\end{equation}
where $\Theta$ represents the parameters to be trained in the neural network. The presence of these parameters to be solved in the neural network greatly enhances the ability of the network to express and extract features. And $\mathcal{G}_\Theta$ seeks for $\Theta^*$ such that $\mathcal{G}_{\Theta^*}\approx \mathcal{G}$ by minimize the loss function $\mathcal{J}\left(\Theta\right)$ defined as follow
\begin{equation}\label{loss}
\mathcal{J}\left(\Theta\right):=\mathbb{E}_{h_1 \sim \mathcal{H}_1}\left[\|\mathcal{G}_{\Theta}(h_1)-h_2 \|^{2}\right].
\end{equation}
It's noted that the relative $L_2$ norm are choose as the $||\cdot||$ in Eq.~(\ref{loss}).

\subsection{The FNO architecture}
 For the input $h_1(x,t)\in\mathcal{R}^{H\times W}$ with given batch size (B), height (H), width (W) and channel ($C_1$),  FNO transforms it to the higher dimension by adding the full connected layer $P$ firstky, i.e $v_0=P(h_1)$.  In the forward problem, the input is mostly chosen as $h_1(x,t_0)\in\mathcal{R}^{H\times W}$ with $t_0$ denoting the initial moment. However, we extend FNO to learn the mapping between the fractional order to the soliton space in the whole spatial-temporal area. That is, the input is chosen as $\epsilon\in\mathcal{R}^{H\times W}$ while the output denotes the solution $h_2(x,t)\in\mathcal{R}^{H\times W}$. On e can also treat $\epsilon$ as the initial condition, i.e. $h_1(x,t_0)=\epsilon$. It is worth mentioning that the position encoding is  also incorporated into the input of the network. Then the fourier layers are applied to extract  the frequency information, and the fourier layer is updated iteratively by
 \begin{equation}\label{update}
v_{t+1}(x)=\sigma\left(W v_{t}(x)+\left(\mathcal{K}(h_1 ; \Theta) v_{t}\right)(x)\right), \quad t=0,1,\cdots,T-1,
\end{equation}
where $\sigma$ denotes the nonlinear activation function which can significantly enhance the ability to express characteristics of the network. And $W$ represents the linear transform to be learned,  $\mathcal{K}(h_1 ; \Theta)$ represents the kernel integral operator (i.e., kernel function) parameterized by the $\Theta$ which is defined as
\begin{equation}\label{Kernal}
\left(\mathcal{K}(h_1 ; \Theta) v_{t}\right)(x):=\int_{D} \kappa(x, y, h_1(x), h_1(y) ; \Theta) v_{t}(y) \mathrm{d} y,
\end{equation}
then supposing $\kappa(x,y;\Theta)=\kappa(x-y;\Theta)$ and applying convolution theorem, $\mathcal{K}$ can be implemented  as
\begin{equation}
\left(\mathcal{K}(\Theta) \nu_{k}\right)(x)=\mathcal{F}^{-1}\left(R_{\Theta} \cdot\left(\mathcal{F}_{\nu_{k}}\right)\right)(x),
\end{equation}
with  $R_{\Theta}$ representing the Fourier transform of a periodic function $\kappa$, and can be  learned via the scalar convolution kernel, $\mathcal{F}$ and $\mathcal{F}^{-1}$ denote the Fourier transform and its inverse, respectively:
 \bee\nonumber
 \mathcal{F}[f(x)](k)=\int_{D}f(x)e^{-2\pi ikx}dx,\quad \mathcal{F}^{-1}[\mathcal{F}[f](k)]
 =\int_{D}\mathcal{F}[f(x)](k)e^{2\pi ikx}dk.
 \ene
 And in the realization, $\mathcal{F}, \mathcal{F}^{-1}$ can be obtained by the fast Fourier transform (FFT) of finite modes. It is because of the presence of Fourier layers that the ability of FNO to extract features is much improved compared to traditional neural networks~\cite{Xu19,Luo19}. More complex structures in physical space are extracted and expressed in frequency domain space by Fourier layers. Lastly, two fully connected layers $Q$ are applied to transform the channel for the problem in hand. Thus, the output for the FNO can be expressed in the form of
\begin{equation}\label{OUTPUT}
  \mathcal{G}_{\Theta}(h_1(:))=Q\circ F_T\circ\cdots\circ F_1\circ P(h_1).
\end{equation}
where $P$ encodes the lower dimension function into higher dimensional space and $Q$ decodes the higher dimension function back to the lower dimensional space. The final dimension of $\mathcal{G}_{\Theta}(h_1(:))$ is  $B\times H\times W\times C_2$. In the following discussion, the Gelu function $\sigma(x)=x\Phi(x)$ is chosen as the nonlinear activation function, with $\Phi(x)$ denoting the  distribution function of the standard normal distribution. The Adam optimizer with the initial learning rate $\eta = 0.001$ halved every 100 epochs in the total 500 epochs. The loss function defined in Eq.~(\ref{loss}) can be approximated by
\begin{equation}\label{Loss2}
\mathcal{J}\left(\Theta\right)=\frac{1}{N}\sum_{j=1}^{N}\frac{\|\mathcal{G}_{\Theta}(h_1^j)-h_2^j \|_{2}}{\|h_2^j\|_{2}}
\end{equation}
with $\mathcal{G}_{\Theta}(h_1^j),h_2^j$ denote the ground truth  and the prediction soliton of the $j$th samples in train set, respectively.

\subsection{The choice of training data}

In the supervised deep learning, the quality of the data determines the training and prediction performance to a certain extent. One can access the data from the exact solution (if available) or the numerical solutions via the high-accuracy numerical methods (e.g., spectral method, finite element method, etc.). Also, for other problems in real life, the data can be derived from experiments and observations. In the following experiments, the fractional-order index $\epsilon$ are generated from the uniform distribution on $\epsilon\in(0,1)$, thus we can get the input for the neural network. Though the parameter only varies from $(0,1)$, the impact on the solitons in the whole domain is huge which can be found in the  discussions.  If not specified, the total number of samples is 1500, of which 1000 are used for training and 200 for the test set. And the batch size is chosen as 20 for mini-batch stochastic gradient descent method.

\section{Data-driven solitons in fractional integrable nonlinear equations}

In this section, we consider the fractional one-solitons in some fractional integrable nonlinear wave equations via FNO. After training, the network is fed with the fractional-order index $\epsilon\in (0, 1)$ and the corresponding solitons are output. And the fKdV, fNLS, fmKdV,
and fsineG  equations are learned, respectively. If not specified, the  spatio-temporal domain $(x, t)\in [-10,10]\times[-1,1]$ are discreted into $256\times100$ grids for the finite collections of output.

\subsection{Fractional KdV equation}

we study the mapping between the fraction-order index $\epsilon$ and the soliton of fKdV equation in this subsection. The one-soliton of the fKdV can be written as~\cite{fraction1}
\begin{equation}\label{fKdVs}
u(x, t)=2 \kappa^{2} \operatorname{sech}^{2}\left(\kappa\left[x-(4 \kappa^{2})^{1+\varepsilon} t\right]\right),
\end{equation}
where $i\kappa,\, \kappa\in\mathbb{R}$ denotes one discrete spectrum in the IST. The one-soliton is a right-going travelling wave with the wave velocity $(4 \kappa^{2})^{1+\varepsilon}$, which is a $2(4 \kappa^{2})^{\varepsilon}$ times more than the amplitude.

We choose $\kappa=1$ for the ground truth data. One can observe that $\epsilon$ only appears in the $t$-part, whose main reason is that the spatial part of the Lax pair remains unchanged, while the fractional order derivative only appears in the dispersion relation (time evolution).  Fourier neural operator aims to learn the mapping between the fractional order  to the soliton space, $\mathcal{G}:\mathcal{R}(0,1)\rightarrow L^{2}\left(x\in[-10, 10];\, t\in[-1,1];\, \epsilon\in(0, 1)\right)$ defined by $\epsilon\in(0,1)\rightarrow u|_{[-10,10]\times[-1,1]}$. We can add the channel for the input of parameter $\epsilon$ ($u(x,0)=\epsilon$).

After training FNO with 500 epochs, the parameters $\Theta$ in the neural network can be obtained. It can be seen that the $L_2$ error on the train/test sets decays  rapidly in the first 300 epochs and then decreases at a very slow rate in the last 200 epochs from Fig.~\ref{Q1DM}(a).  We can also observe   that there exist sharp downward trends at 100, 200, 300 epochs from the enlarged plot in Fig.~\ref{Q1DM}(a), which is mainly due to the change in learning rate here. The errors finally reach \{1.224e-3, 1.229e-3\}. Moreover, the relative $L_2$ error on test sets are also recorded to verify the generalization ability of FNO. Fig.~\ref{Q1DM}(b) displays the $L_2$ error between the predicted soliton and the corresponding ground truth.  It is worth pointing out that the unit of error here is $10^{-4}$ (similar hereinafter). Once the network is trained, FNO can quickly give the corresponding soliton solutions for $\epsilon$ on the test set. It can be concluded that FNO performs well in the test sets even though the $\epsilon$ has an strong impact on the soliton.

\begin{figure*}[!t]
    \centering
\vspace{-0.15in}
  {\scalebox{0.56}[0.56]{\includegraphics{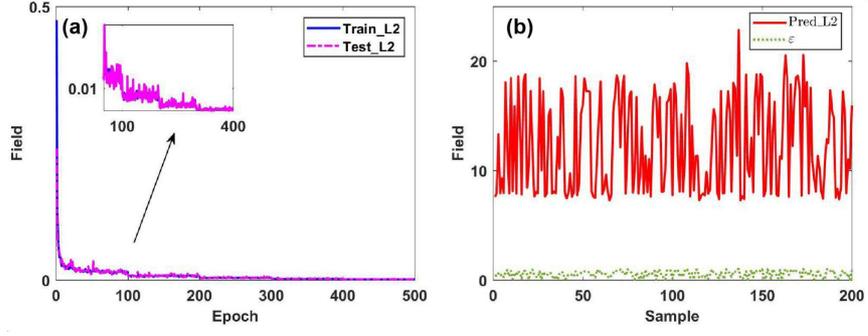}}}\hspace{-0.35in}
\vspace{0.1in}
\caption{The training and testing progresses. (a) The   relative $L_2$ error on training (loss) and test sets versus epoch. (b) The predicted $L_2$ error ($10^{-4}$) and the sampled points of parameter $\epsilon\in (0, 1)$.}
  \label{Q1DM}
\end{figure*}
\begin{figure*}[!t]
    \centering
\vspace{-0.15in}
  {\scalebox{0.7}[0.7]{\includegraphics{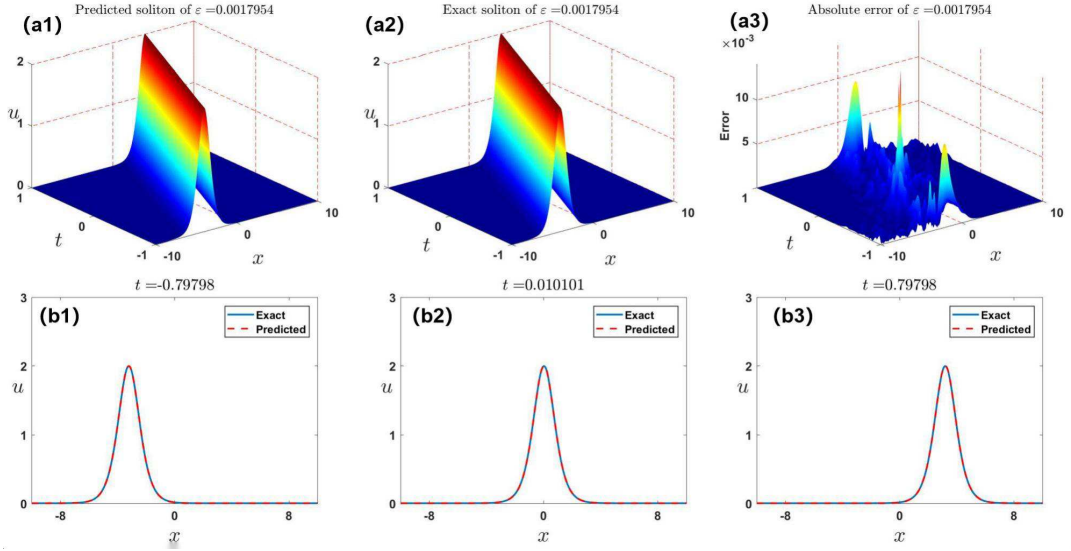}}}\hspace{-0.35in}
\vspace{0.1in}
\caption{The data-driven soliton  of the fKdV equation (\ref{fKdV}) via the FNO method for $\epsilon=0.0017954$. The 3D profiles of the predicted soliton (a1) and exact soliton (a2); (a3) The absolute error between the exact and predicted solutions; (b1-b3) The comparisons between the reference and predicted solitons at  three $t$-snapshots $ t=-0.79798,0.010101$ and $0.79798$, respectively. }
  \label{Q1minM}
\end{figure*}

To better visualize the data-driven solitons of fKdV, we choose the max and min of $\epsilon$ in the test set containing 200 sample points generated from the uniform distribution on $(0,1)$ such that we can see the impact of $\epsilon$ on the soliton dynamics and the FNO's performance.  For the case of $\epsilon=0.0017954$ (i.e., the min of $\epsilon$ in the test set), the dynamic behavior of the learned solution, the exact solution, and the absolute error between them are demonstrated in Figs.~\ref{Q1minM}(a1-a3). It can be seen that the predicted soliton excellently agrees with the exact one. The absolute error is mostly below $0.01$. Besides, we also choose three $t-$snapshots exhibited in Figs.~\ref{Q1minM}(b1-b3) to exhibit the FNO's effects. With the increase of time, a tendency of movement of the soliton can be seen. And the FNO's predicted soliton matches well with the exact solution.  We then choose the max $\epsilon=0.9959$ in the test set to present the influence of $\epsilon$ on the soliton behavior. As it can be seen that
the soliton for $\epsilon=0.9959$ propagates more quickly than one for $\epsilon=0.0017954$ in this case from  Figs.~\ref{Q1maxM}(a1-a3), which can also be obtained from Eq.~(\ref{fKdVs}). The max absolute error only reaches about 0.02, which implies the powerful generalization ability of FNO. And the predicted soliton also displays the high accuracy from Figs.~\ref{Q1maxM}(b1-b3) in three $t$-snapshots.

On the basis of the above simulations, we can draw conclusion that for the input $\epsilon$, FNO is able to extract the key features with only 1000 samples. The fully connected layers $P$ and $Q$, as well as the Fourier layers all play the critical role in the network. More importantly, the network is able to give the output $u\in\mathcal{H}_2$ in the whole spatial-temporal domain, which differs from the forward problems of PDEs in PINNs.  Surprisingly,  the solution learned by FNO also achieves SOTA in terms of accuracy.

\begin{figure*}[!t]
    \centering
\vspace{-0.15in}
  {\scalebox{0.7}[0.7]{\includegraphics{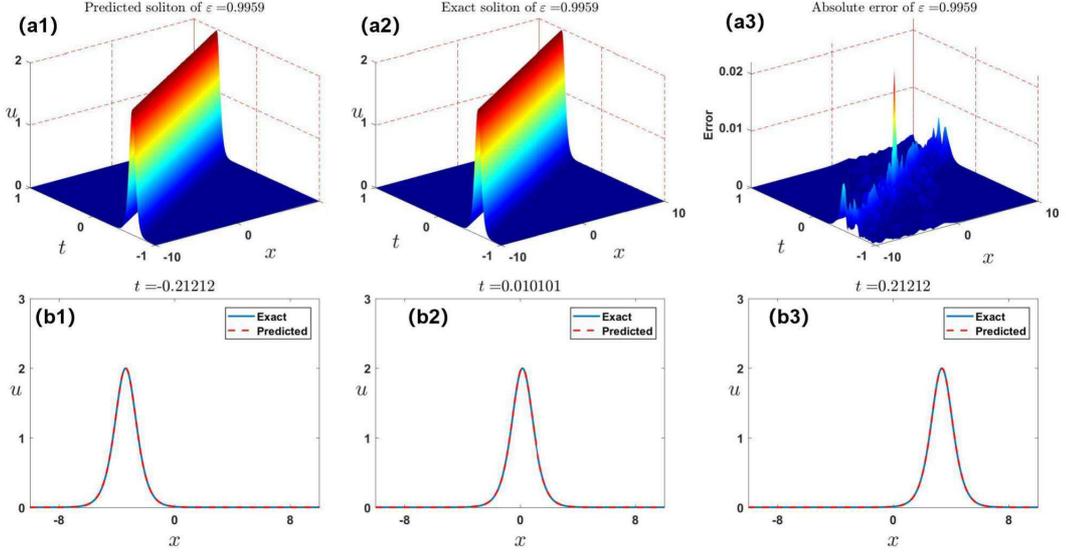}}}\hspace{-0.35in}
\vspace{0.1in}
\caption{The data-driven soliton  of the fKdV equation (\ref{fKdV}) via the FNO method for $\epsilon=0.9959$. The 3D profiles of
the predicted soliton (a1) and exact soliton (a2); (a3) The absolute error between the exact and predicted solitons; (b1-b3) The comparisons between the reference and predicted solitons at three $t$-snapshots $ t=-0.21212,0.010101$ and $0.21212$, respectively. }
  \label{Q1maxM}
\end{figure*}

\subsection{Fractional NLS equation}

Here we consider the data-driven soliton of the focusing fNLS equation (\ref{fNLS}) in the form\cite{fraction1}
\begin{equation}\label{fNLSs}
\psi(x, t)=2 \eta \operatorname{sech}\left(X_{\epsilon}(x, t)\right)e^{i[-2\xi x+4\left(\xi^{2}-\eta^{2}\right)\left|2k\right|^{\epsilon} t]} ,
\end{equation}
where $X_{\epsilon}(x, t)=2 \eta\left(x-4 \xi\left|2 k\right|^{\epsilon} t\right)$ and $k=\xi+i \eta,\,\, (\xi,\, \eta>0\in\mathbb{R})$ denotes the discrete spectrum. This is a right (left)-going travelling wave with the wave velocity
$2^{2+\epsilon}\xi(\xi^2+\eta^2)^{\epsilon/2}$ and $\xi>0\, (\xi<0)$. The phase velocity of the envelope soliton is
$2^{1+\epsilon}(\xi^2-\eta^2)(\xi^2+\eta^2)^{\epsilon/2}/\xi$.

 We choose $\eta=0.5,\, \xi=1.5$ in the following discussions. And it's easy to verify that $|\psi(x,t)|=2|\eta| \operatorname{sech} \left(X_{\epsilon}(x, t)\right).$ FNO here aims to learn the mapping between the fractional order space $\epsilon$ to the soltion space which can be defined by $\epsilon\in(0,1)\rightarrow \psi|_{[-10,10]\times[-1,1]}$. It's noted that the output of the neural network is complex-valued, thus we should rewrite it as the real and imaginary parts. And it can be implemented through adding another channel. The  parameter $\Theta$  in the neural network  can be trained via  mini-batch gradient descent using the Adam optimizer with decay learning rate. The $L_2$ error on the training  and test sets versus epoch is displayed in Fig.~\ref{Q2DM}(a). One can observe that the $L_2$ error decays quickly in the first 300 epoch then  decline slowly. Similar with the former case, the error  decreases sharply at 100,\, 200,\, 300,\, 400 epochs, which is mainly due to the change in learning rate. The error reaches 2.257e-3, 2.273e-3, respectively. Moreover, we also consider the performance of FNO on the tese sets (see Fig.~\ref{Q2DM}(b)). It can be seen that the relative $L_2$ error is mostly below 6.0e-4, which demonstrates the powerful generalization ability of FNO.

We also display the data-driven soliton of fNLS for the visualization. For $\epsilon=0.012226$, the predicted soliton, exact soliton and the absolute error between them are exhibited in Fig.~\ref{Q2minM}.  It can be concluded that the learned soliton via FNO matches well with the exact solution from Figs.~\ref{Q2minM}(a1-a3). The absolute error is mostly below 0.01. To better observe the dynamic behaviour, three  different $t$-snapshots are presented in Figs.~\ref{Q2minM}(b1-b3). One can see the trend of movement from Figs.~\ref{Q2minM}(b1-b3). Moreover, the predicted soliton achieves an excellent agreement with the ground truth. We then choose $\epsilon=0.99893$ to demonstrate the influence of $\epsilon$ on the soliton dynamic behaviours and FNO's performance. Similar with the case of fKdV,  the soliton for $\epsilon=0.99893$ propagates more quickly than one for $\epsilon=0.012226$ from Figs.~\ref{Q2maxM}(a1-a3). The data-driven soliton matches well with the ground truth and the max absolute error is only 0.02.  And the predicted soliton also displays the high accuracy from Figs.~\ref{Q2maxM}(b1-b3) in three $t$-snapshots.

It can be concluded that FNO performs well in the mapping between the fractional order space and the soliton space from the above results. For the new instance $\epsilon$ in the test sets, FNO can give the soliton in the whole spatio-temporal domain quickly. Moreover, the predicted solution  also matches well with the exact solution.

\begin{figure*}[!t]
    \centering
\vspace{-0.15in}
  {\scalebox{0.56}[0.56]{\includegraphics{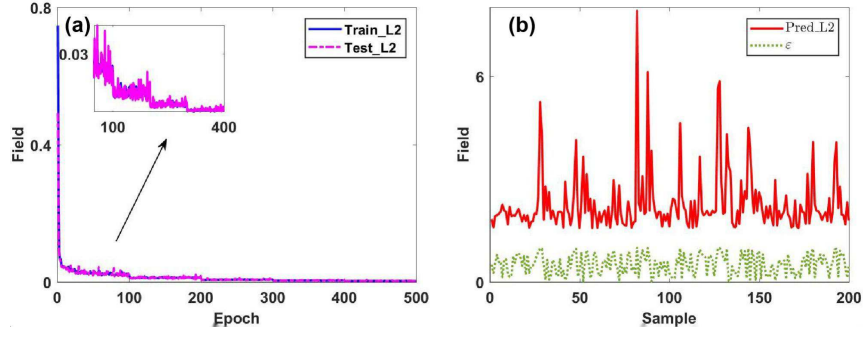}}}\hspace{-0.35in}
\vspace{0.1in}
\caption{The training and testing progresses. (a) The   relative $L_2$ error on training (loss) and test sets versus epoch. (b) The predicted $L_2$ error ($10^{-4}$) and the sampled points of  parameter $\epsilon\in (0, 1)$.}
  \label{Q2DM}
\end{figure*}

\begin{figure*}[!t]
    \centering
\vspace{-0.1in}
  {\scalebox{0.7}[0.7]{\includegraphics{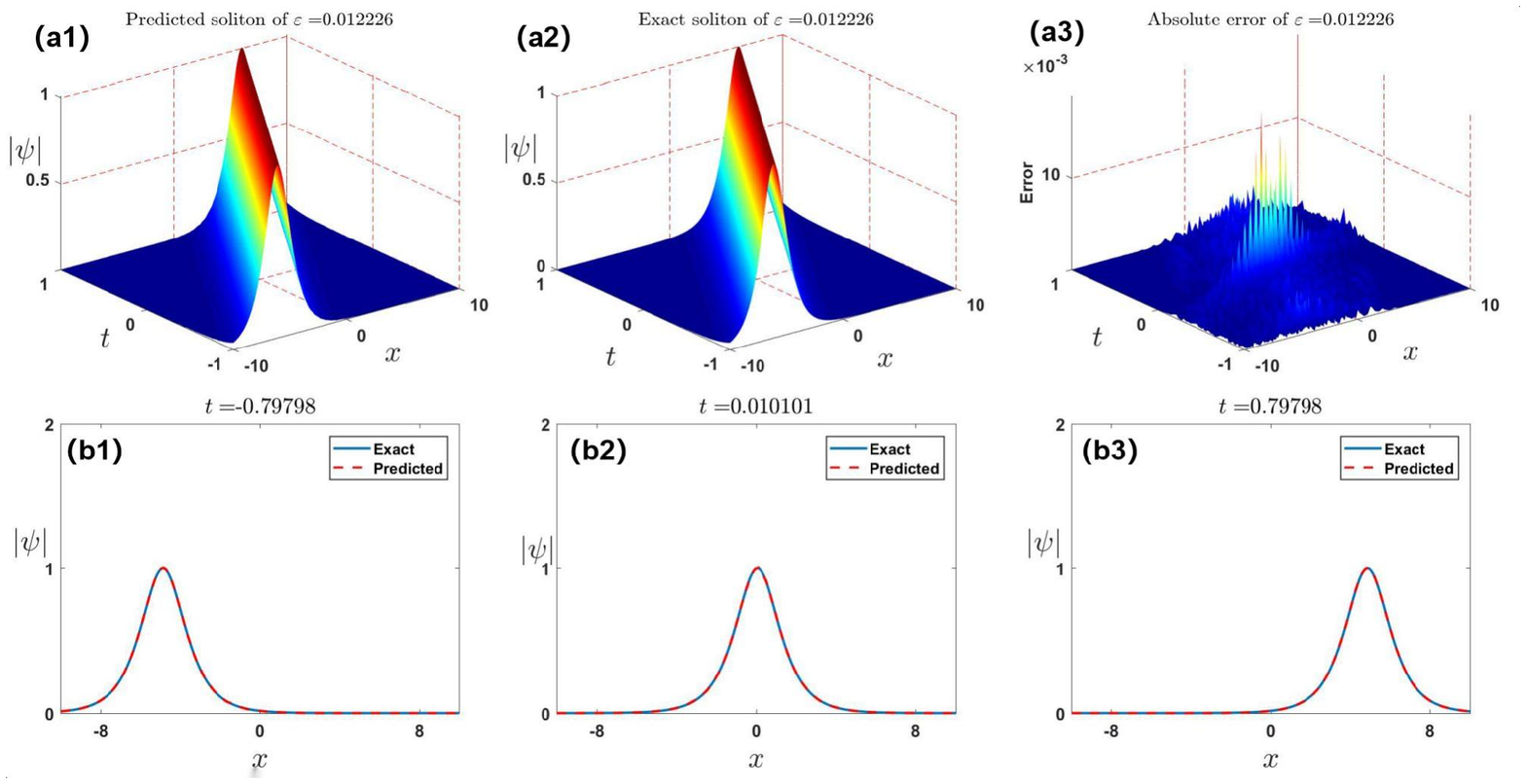}}}\hspace{-0.35in}
\vspace{0.1in}
\caption{The data-driven soliton  of the fNLS equation (\ref{fNLS}) via the FNO method for $\epsilon=0.012226$. The 3D profiles of the predicted soliton (a1) and exact soliton (a2); (a3) The absolute error between the exact and predicted solitons; (b1-b3) The comparisons between the reference  and predicted solitons at three $t$-snapshots $ t=-0.79798,0.010101$ and $0.79798$, respectively. }
  \label{Q2minM}
\end{figure*}

\begin{figure*}[!t]
    \centering
\vspace{-0.15in}
  {\scalebox{0.7}[0.7]{\includegraphics{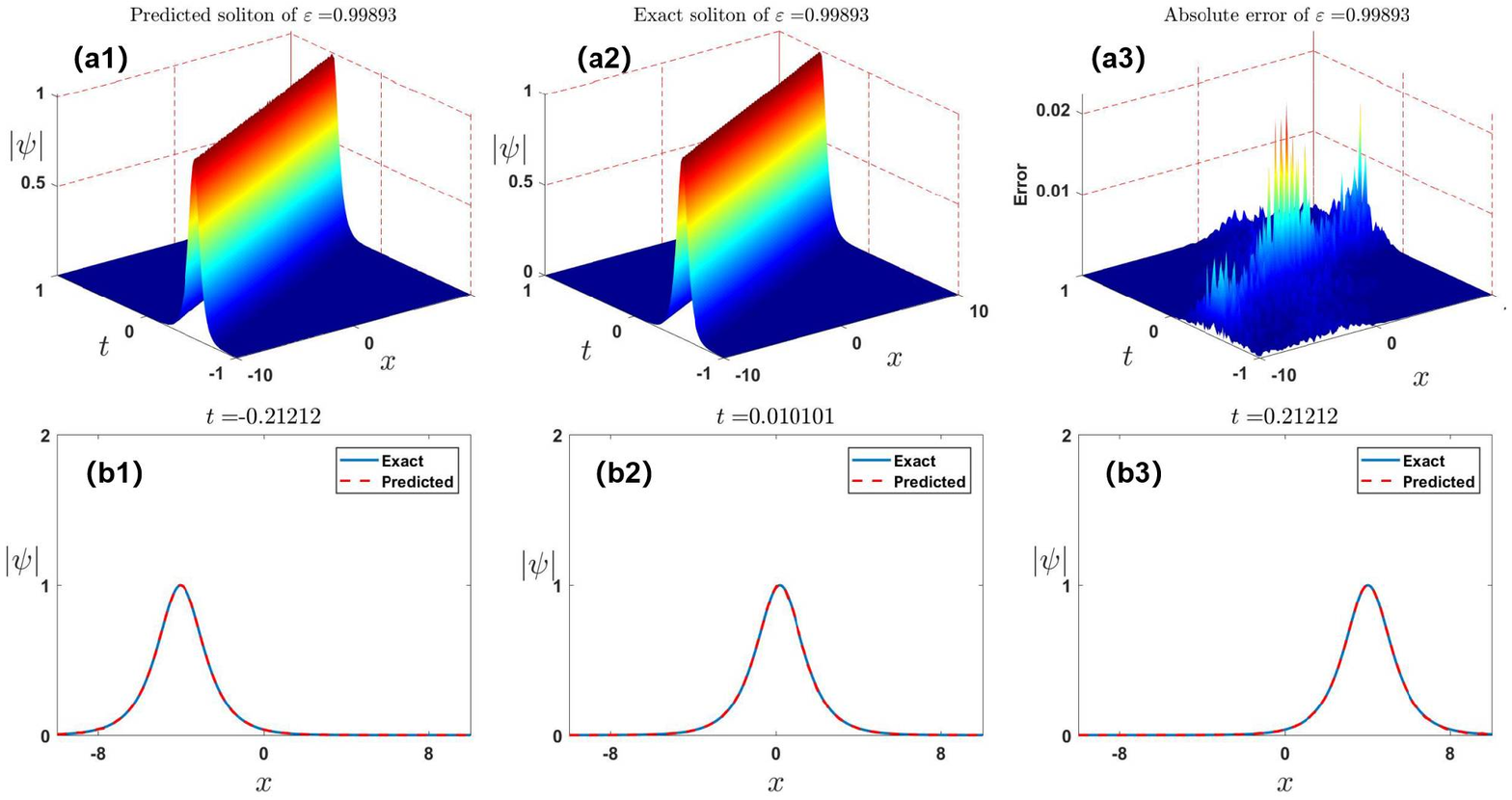}}}\hspace{-0.35in}
\vspace{0.1in}
\caption{The data-driven soliton  of the fNLS equation (\ref{fNLS}) via the FNO method for $\epsilon=0.99893$. The 3D profiles of the predicted soliton (a1) and exact soliton (a2); (a3) The absolute error between the exact and predicted solitons; (b1-b3) The comparisons between the reference and predicted solitons at three $t$-snapshots $ t=-0.21212,0.010101$ and $0.21212$, respectively. }
  \label{Q2maxM}
\end{figure*}

\subsection{Fractional mKdV equation}
\begin{figure*}[!t]
    \centering
\vspace{-0.15in}
  {\scalebox{0.56}[0.56]{\includegraphics{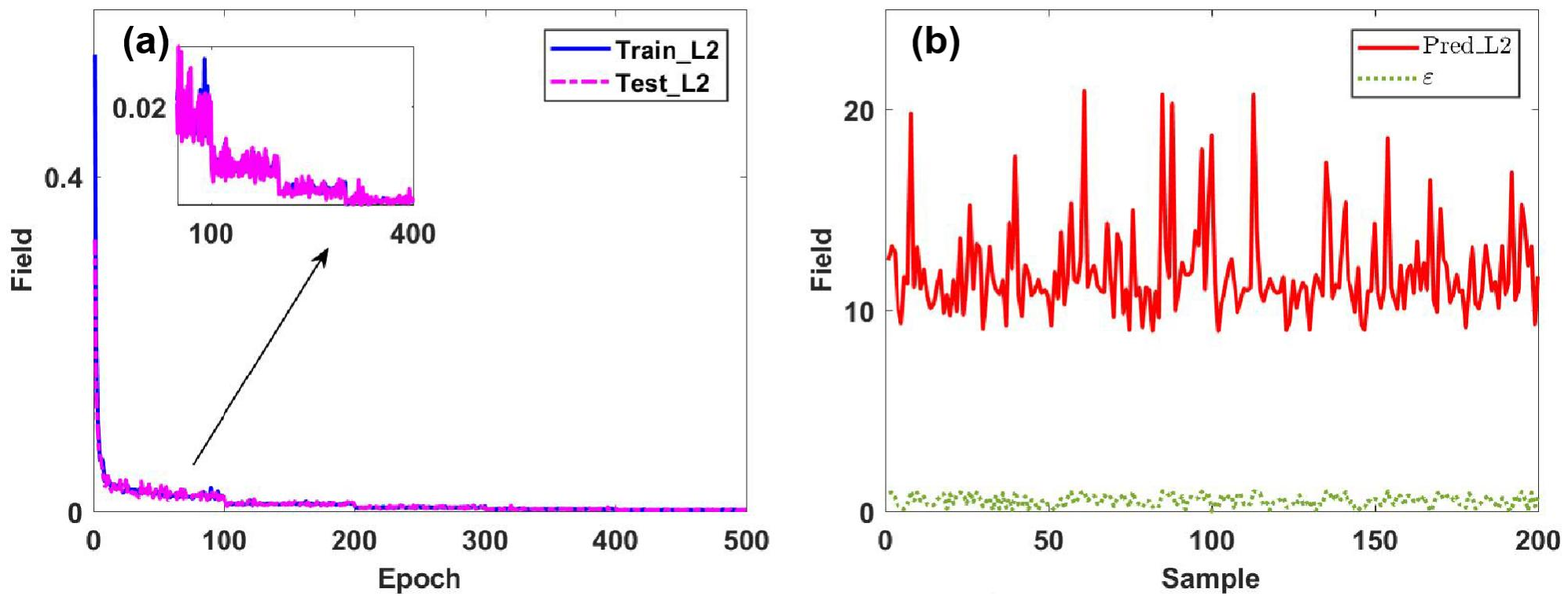}}}\hspace{-0.35in}
\vspace{0.1in}
\caption{The training and testing progresses. (a) The   relative $L_2$ error on training (loss) and test sets versus epoch. (b) The predicted $L_2$ error ($10^{-4}$) and the sampled points of parameter $\epsilon \in (0, 1)$.}
  \label{Q3DM}
\end{figure*}

\begin{figure*}[!t]
    \centering
\vspace{-0.15in}
  {\scalebox{0.75}[0.75]{\includegraphics{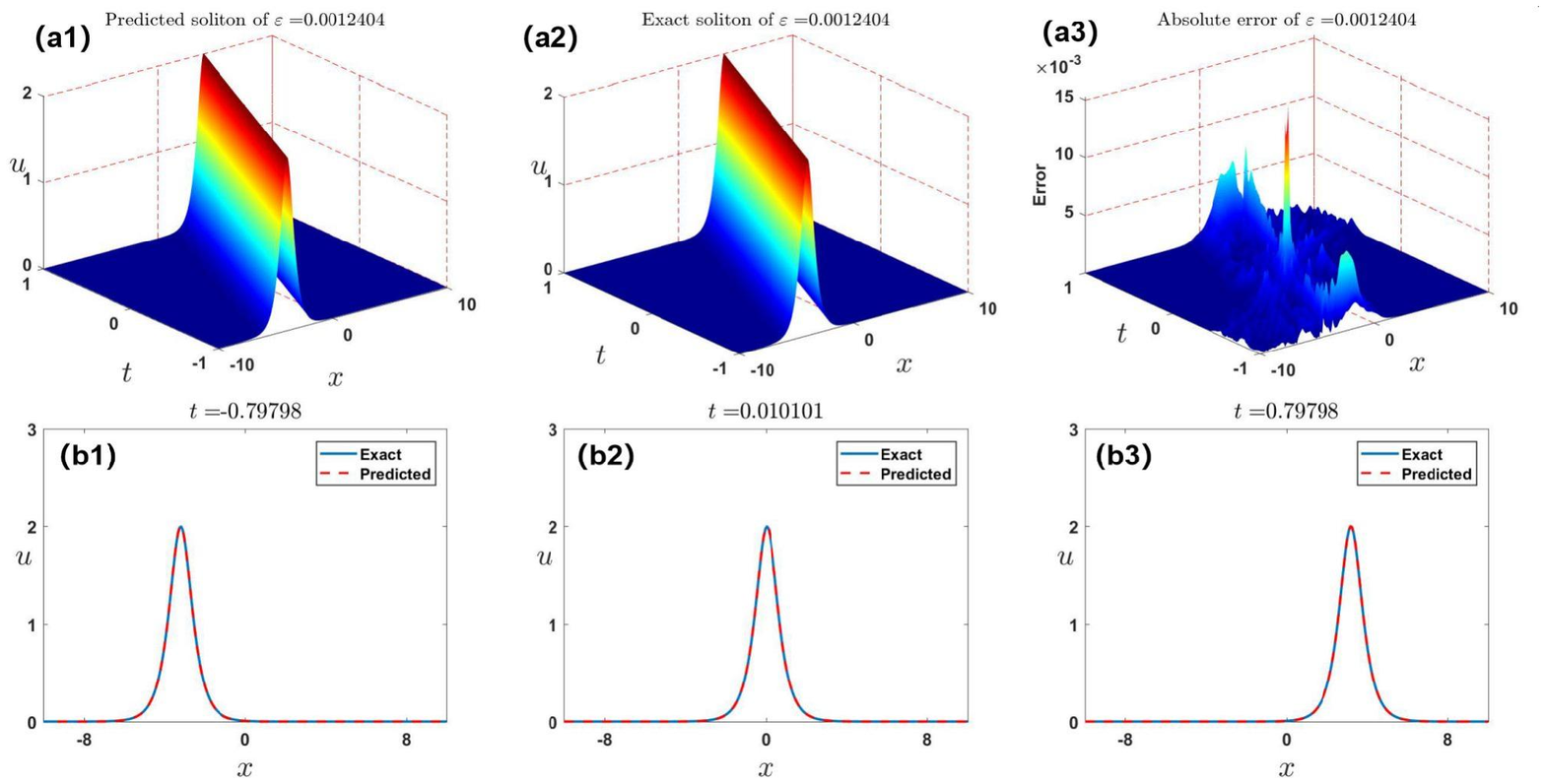}}}\hspace{-0.35in}
\vspace{0.1in}
\caption{The data-driven soliton  of the fmKdV equation via the FNO method for $\epsilon=0.0012404$. The 3D profiles of the predicted soliton (a1) and exact soliton (a2); (a3) The absolute error between the exact and predicted solitons; (b1-b3) The comparisons between the reference  and predicted solitons at three $t$-snapshots $ t=-0.79798,0.010101$ and $0.79798$, respectively. }
  \label{Q3minM}
\end{figure*}

\begin{figure*}[!t]
    \centering
\vspace{-0.15in}
  {\scalebox{0.7}[0.7]{\includegraphics{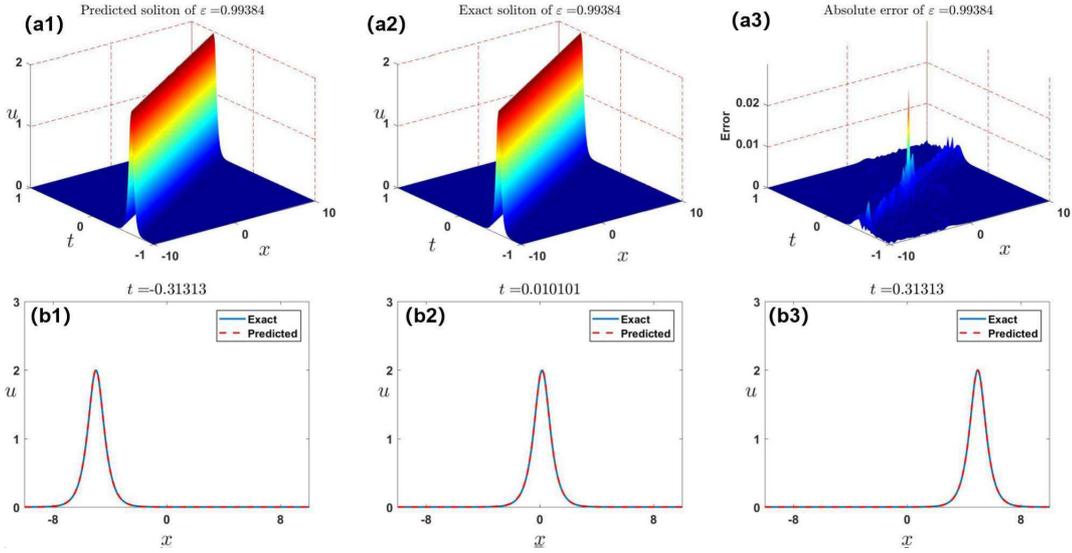}}}\hspace{-0.35in}
\vspace{0.1in}
\caption{The data-driven soliton  of the fmKdV equation  via the FNO method for $\epsilon=0.99384$. The 3D profiles of the predicted soliton (a1) and exact soliton (a2); (a3) The absolute error between the exact and predicted solitons; (b1-b3) The comparisons between the reference and predicted solitons at three $t$-snapshots $ t=-0.31313, 0.010101$ and $0.31313$, respectively. }
  \label{Q3maxM}
\end{figure*}
In this subsection, we mainly consider the data-driven soliton of the focusing fmKdV equation in the form~\cite{fraction2}
\begin{equation}\label{fMKDVs}
u(x, t)=2 \eta \operatorname{sech}\left(2 \eta x-(2 \eta)^{3+2 \epsilon} t\right),
\end{equation}
where $i\eta,\, \eta\in\mathbb{R}^+$ denotes the discrete spectrum in the  inverse scattering transform. This is a right-going travelling wave with the wave velocity $(2 \eta)^{2+2 \epsilon}$, which is $(2 \eta)^{1+2 \epsilon}$ times more than the amplitude.

We take $\eta=1$ for the ground truth data.  Fourier neural operator aims to learn the mapping between the fraction order  to the soliton space, $\mathcal{G}:\mathcal{R}(0,1)\rightarrow L^{2}\left(x\in[-10, 10];\, t\in[-1,1];\, \epsilon\in (0, 1)\right)$ defined by $\epsilon\in(0,1)\rightarrow u|_{[-10,10]\times[-1,1]}$. We can add the channel for the input of $\epsilon$.
After training FNO with 500 epochs, the parameters $\Theta$ in the neural network can be obtained. It can be seen that the $L_2$ error on the train/test sets decays  rapidly in the first 300 epochs and then decreases at a very slow rate in the last 200 epochs from Fig.~\ref{Q3DM}(a).  We can also observe   that there exist sharp downward trends at 100, 200, 300 epochs from the enlarged plot in Fig.~\ref{Q1DM}(a), which is mainly due to the change in learning rate here. The errors finally reach \{1.187e-3, 1.519e-3\}. Moreover, the relative $L_2$ error on test sets are also recorded to verify the generalization ability of FNO. Fig.~\ref{Q3DM}(b) displays the $L_2$ error between the predicted soliton and the corresponding ground truth.   Once the network is trained, FNO can quickly give the corresponding soliton solutions for $\epsilon$ on the test set. It can be concluded that FNO performs well in the test sets even though the $\epsilon$ has an strong impact on the soliton solution.

To better visualize the data-driven solitons of fmKdV, we choose the max and min of $\epsilon$ in the test set such that we can see the impact of $\epsilon$ on the soliton dynamics and the FNO's performance.  For the case of $\epsilon=0.0012404$, the dynamic behavior of the learned solution, the exact solution, and the absolute error between them are demonstrated in Figs.~\ref{Q3minM}(a1-a3), which display that the predicted soliton excellently agrees with the exact solution. The absolute error is mostly below $0.01$. Besides, we also choose three $t-$snapshots exhibited in Figs.~\ref{Q3minM}(b1-b3) to exhibit the FNO's effects. With the increase of time, a tendency of movement of the soliton can be seen. And the FNO's predicted soliton matches well with the exact solution.  We then choose $\epsilon=0.99384$ in the test sets to present the influence of $\epsilon$ on the soliton behaviours. As it can be seen that the  soliton for $\epsilon=0.99384$ propagates more quickly than one for
$\epsilon=0.0012404$ from  Figs.~\ref{Q3maxM}(a1-a3), which can also obtain  from Eq.~(\ref{fMKDVs}). The max absolute error only reaches about 0.02, which exhibit the powerful generalization ability of FNO. And the predicted soliton also displays the high accuracy from Figs.~\ref{Q3maxM}(b1-b3) in three $t$-snapshots.

The above results reveal that FNO is a powerful tool in extracting data features. With only 1000 samples, FNO learns the mapping relationships between data and performs very well on the test set. Surprisingly, the solutions learned by FNO also achieve SOTA in terms of accuracy.

\begin{figure*}[!t]
    \centering
\vspace{-0.15in}
  {\scalebox{0.56}[0.56]{\includegraphics{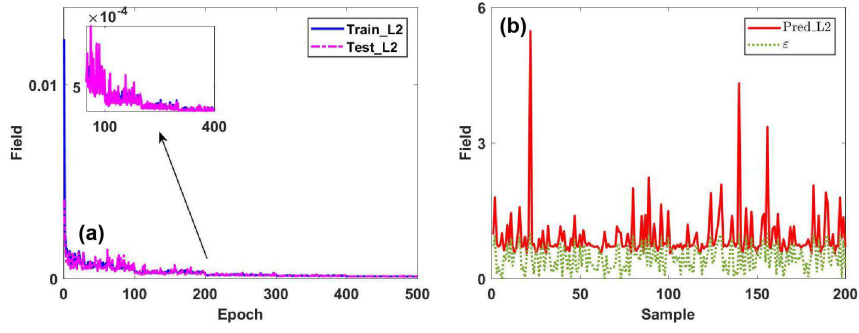}}}\hspace{-0.35in}
\vspace{0.1in}
\caption{The training and testing progresses. (a) The   relative $L_2$ error on training (loss) and test sets versus epoch. (b) The predicted $L_2$ error ($10^{-4}$) and the sampled points of parameter $\epsilon\in (0, 1)$.}
  \label{Q4DM}
\end{figure*}

\begin{figure*}[!t]
    \centering
\vspace{-0.1in}
  {\scalebox{0.7}[0.7]{\includegraphics{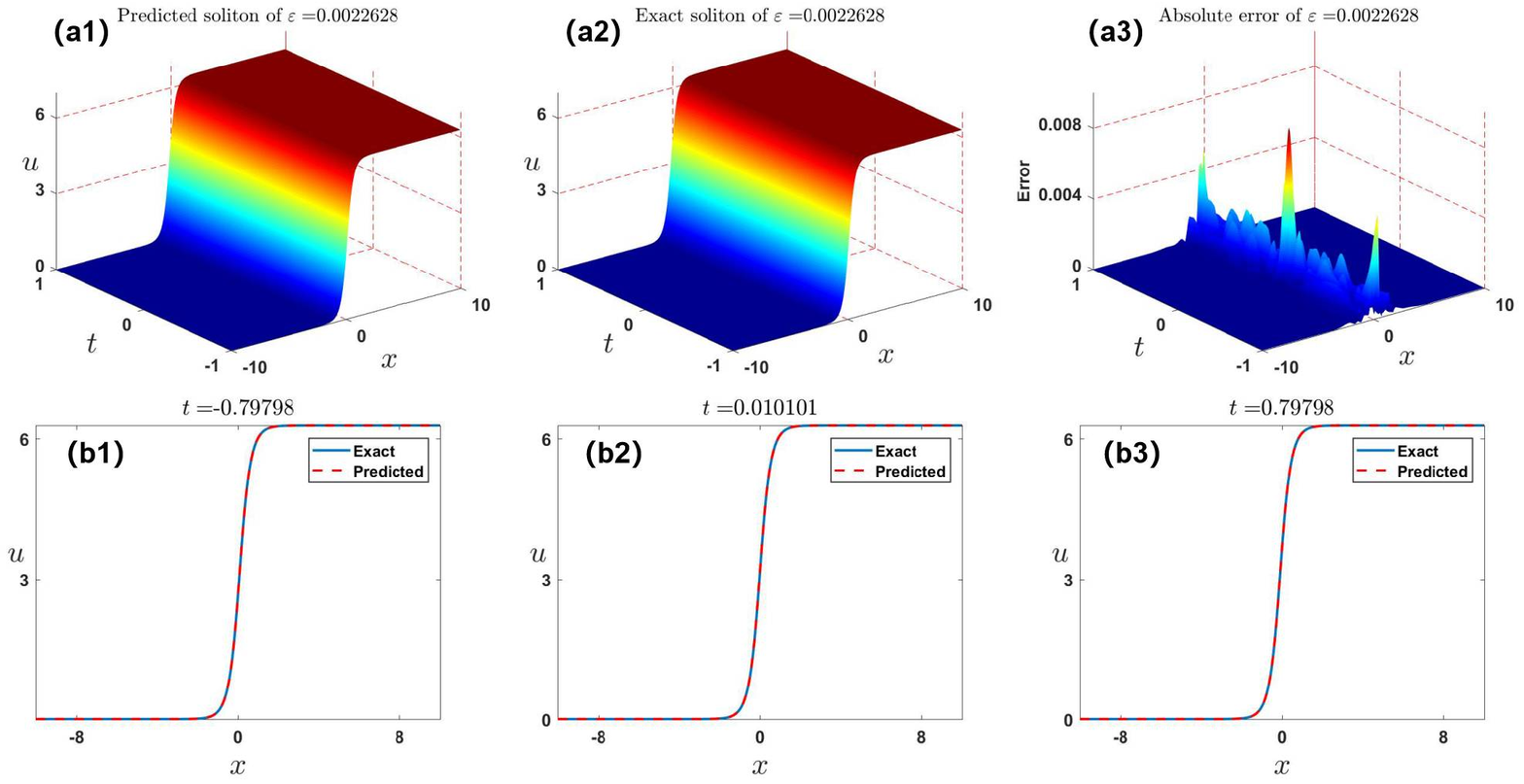}}}\hspace{-0.35in}
\vspace{0.1in}
\caption{The data-driven soliton  of the fsineG equation  via the FNO method for $\epsilon=0.0022628$. The 3D profiles of the predicted soliton (a1) and exact soliton (a2); (a3) The absolute error between the exact and predicted solitons; (b1-b3) The comparisons between the reference and predicted solitons at three $t$-snapshots $ t=-0.79798,0.010101$ and $0.79798$, respectively. }
  \label{Q4minM}
\end{figure*}

\begin{figure*}[!t]
    \centering
\vspace{-0.15in}
  {\scalebox{0.7}[0.7]{\includegraphics{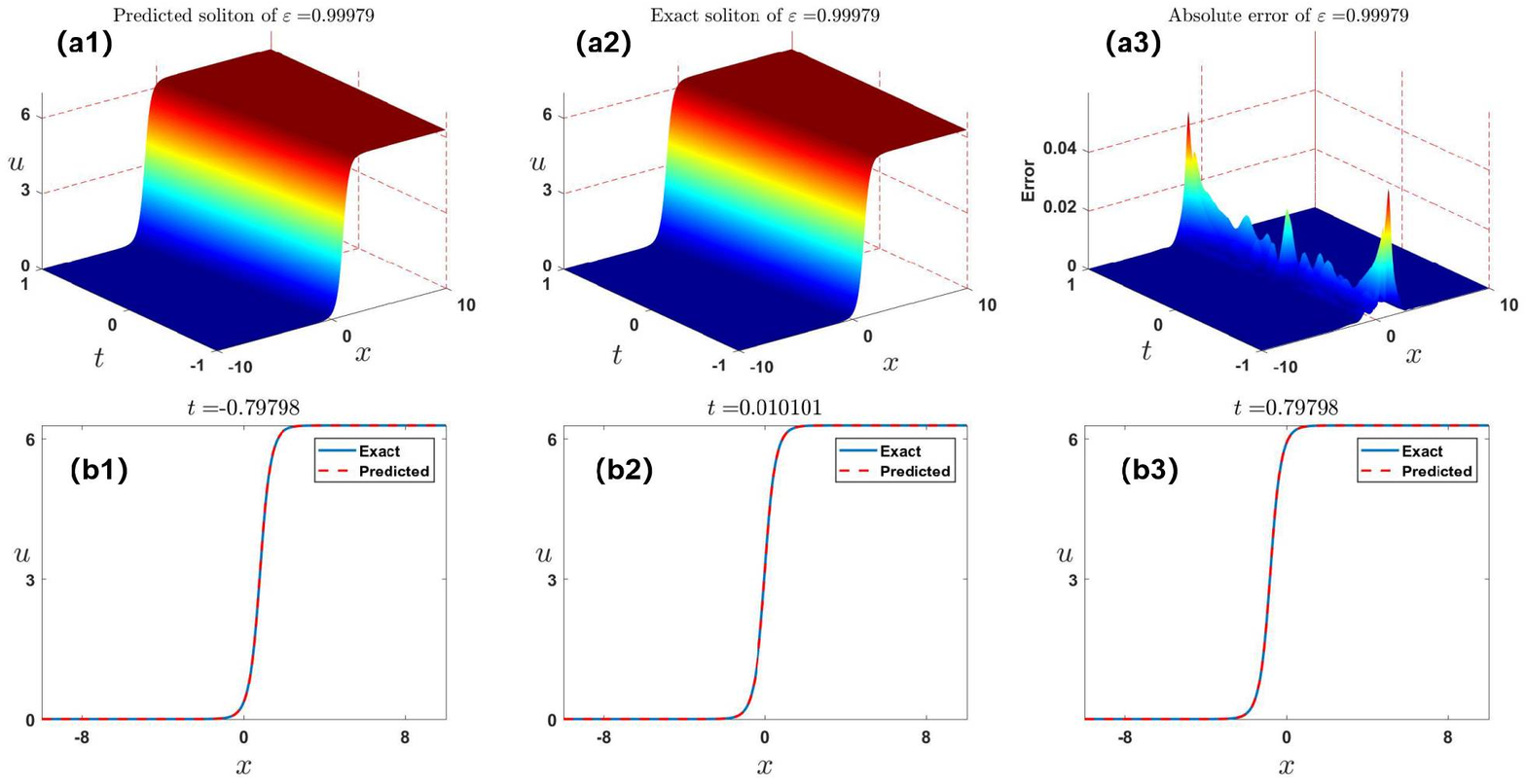}}}\hspace{-0.35in}
\vspace{0.1in}
\caption{The data-driven soliton  of the fsineG equation  via the FNO method for $\epsilon=0.99799$. The 3D profiles of the predicted soliton (a1) and exact soliton (a2); (a3) The absolute error between the exact and predicted solutions; (b1-b3) The comparisons between the reference and predicted solitons at three $t$-snapshots $ t=-0.79798,0.010101$ and $0.79798$, respectively. }
  \label{Q4maxM}
\end{figure*}

\subsection{Fractional sine-Gordon equation}
In this subsection, we mainly consider the data-driven kink soliton of fsineG in the form~\cite{fraction2}
\begin{equation}\label{fSineGs}
u(x, t)=2\arctan \sinh \left(2 \eta x+(2 \eta)^{2 \epsilon-1} t\right)+\pi,
\end{equation}
where $i\eta,\, (\eta\in\mathbb{R}^+)$ denotes the discrete spectrum in the  inverse scattering transform (Notice that the solution is revised). This is a left-going travelling wave with the velocity $(2 \eta)^{2 \epsilon-2}$.

We choose $\eta=3/2$ for the ground truth data.  FNO aims to learn the mapping between the fraction order  to the soliton space, $\mathcal{G}:\mathcal{R}(0,1)\rightarrow L^{2}\left(x\in [-10, 10]; \,t\in[-1,1];\, \epsilon\in (0, 1)\right)$ defined by $\epsilon\in(0,1)\rightarrow u|_{[-10,10]\times[-1,1]}$. We can add the channel for the input of $\epsilon$.
After training FNO with 500 epochs, the parameters $\Theta$ in the neural network can be obtained. It can be seen that the $L_2$ error in the train/test set decays  rapidly in the first 200 epochs and then decreases at a very slow rate in the last 300 epochs from Fig.~\ref{Q4DM}(a).  We can also observe   that there exist sharp downward trends at 100, 200, 300 epochs from the enlarged plot in Fig.~\ref{Q4DM}(a), which is mainly due to the change in learning rate here. The errors finally reach \{9.984e-5, 9.260e-5\}. Moreover, the relative $L_2$ error in test set is also recorded to verify the generalization ability of FNO. Fig.~\ref{Q4DM}(b) displays the $L_2$ error between the predicted soliton and the corresponding ground truth.   Once the network is trained, FNO can quickly give the corresponding soliton solutions for $\epsilon$ in the test set. It can be concluded that FNO performs well in the test set even though the $\epsilon$ has an strong impact on the soliton.

To better visualize the data-driven solitons of fsineG, we choose the max and min of $\epsilon$ in the test set such that we can see the impact of $\epsilon$ on the soliton dynamics and the FNO's performance.  For the case of $\epsilon=0.0022628$, the dynamic behavior of the learned solution, the exact solution, and the absolute error between them are demonstrated in Figs.~\ref{Q4minM}(a1-a3), which display that the predicted soliton excellently agrees with the exact solution. The absolute error is mostly below $0.005$. Besides, we also choose three $t-$snapshots exhibited in Figs.~\ref{Q4minM}(b1-b3) to exhibit the FNO's effects. With the increase of time, a tendency of movement of the soliton can be seen. And the FNO's predicted soliton matches well with the exact solution.  We then choose $\epsilon=0.99979$ in the test set to present the influence of $\epsilon$ on the soliton behaviours. As it can be seen that the soliton for $\epsilon=0.99979$ propagates more quickly than one for $\epsilon=0.0022628$ in this case from  Figs.~\ref{Q4maxM}(a1-a3), which can also be obtained  from Eq.~(\ref{fSineGs}). The max absolute error only reaches about 0.04, which exhibit the powerful generalization ability of FNO. And the predicted soliton also displays the high accuracy from Figs.~\ref{Q4maxM}(b1-b3) in three $t$-snapshots.

On the basis of the above simulations, we can draw conclusion that for the input $\epsilon$, FNO is able to extract the key features with only 1000 samples. The fully connected layers $P$ and $Q$, as well as the Fourier layers all play the critical role in the network. More importantly, the network is able to give the output $u\in\mathcal{H}_2$ in the whole spatial-temporal domain which differs from the forward problem of PDE.  Surprisingly,  the solutions learned by FNO also achieve SOTA in terms of accuracy.

\section{ Comparisons of the critical factors in FNO scheme }

In what follows, we mainly consider the influences of the nonlinear activation functions and the depths for the fully connected layer $P$ on the FNO scheme.

\subsection{The influence of nonlinear activation functions }

\begin{table}
\centering
\caption{Comparisons between nonlinear activation function and relative $L_2$ error.}
\vspace{0.05in}
\begin{tabular}{c|cccc}
\hline\hline
\rule{0pt}{13pt}\diagbox{Function}{Error \qquad}               & \qquad Worst                     & \qquad Mean                        &\qquad Best
 \\[1ex] \hline
\rule{0pt}{13pt} Relu$(x)$                      &  \qquad 3.072e-3    &  \qquad 1.423e-3                      &  \qquad 8.532e-4                                      \\ \\
Sigmoid$(x)$                           &  \qquad 1.668e-2    & \qquad 3.433e-3        &  \qquad 1.646e-3             \\ \\
Swish$(x)$                         &  \qquad 2.542e-3             & \qquad 1.211e-3                   &  \qquad 8.587e-4             \\ \\
$x\tanh(x)$                           &  \qquad 2.944e-3              &  \qquad 9.598e-4                 &  \qquad 5.436e-4             \\[1ex]
\hline\hline
\end{tabular}
\label{Com1}
\end{table}

\begin{figure*}[!t]
    \centering
\vspace{-0.1in}
 {\scalebox{0.75}[0.75]{\includegraphics{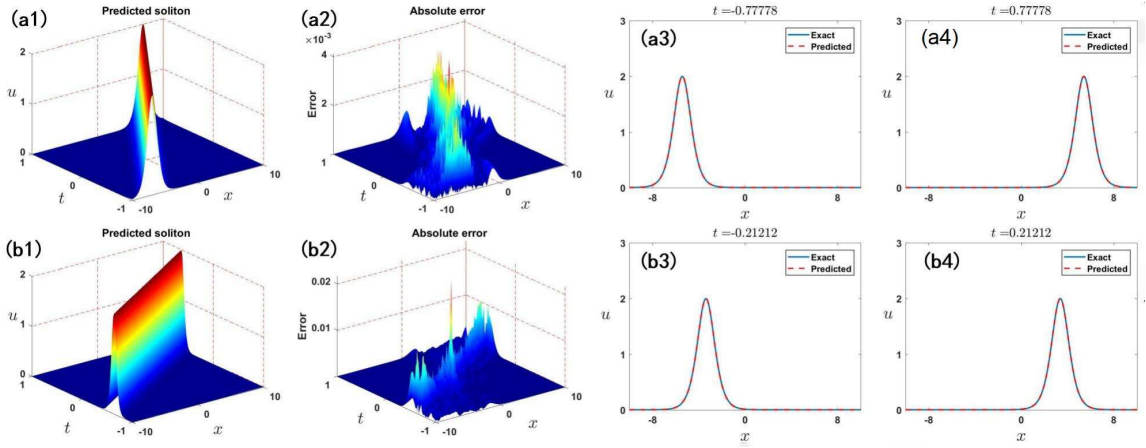}}}\hspace{-0.35in}
\vspace{0.1in}
\caption{(a1)-(a4) best and (b1)-(b4) worst performances of FNO with the Relu$(x)$ activation function.    }
  \label{reluM}
\end{figure*}

\begin{figure*}[!t]
    \centering
\vspace{-0.1in}
 {\scalebox{0.75}[0.75]{\includegraphics{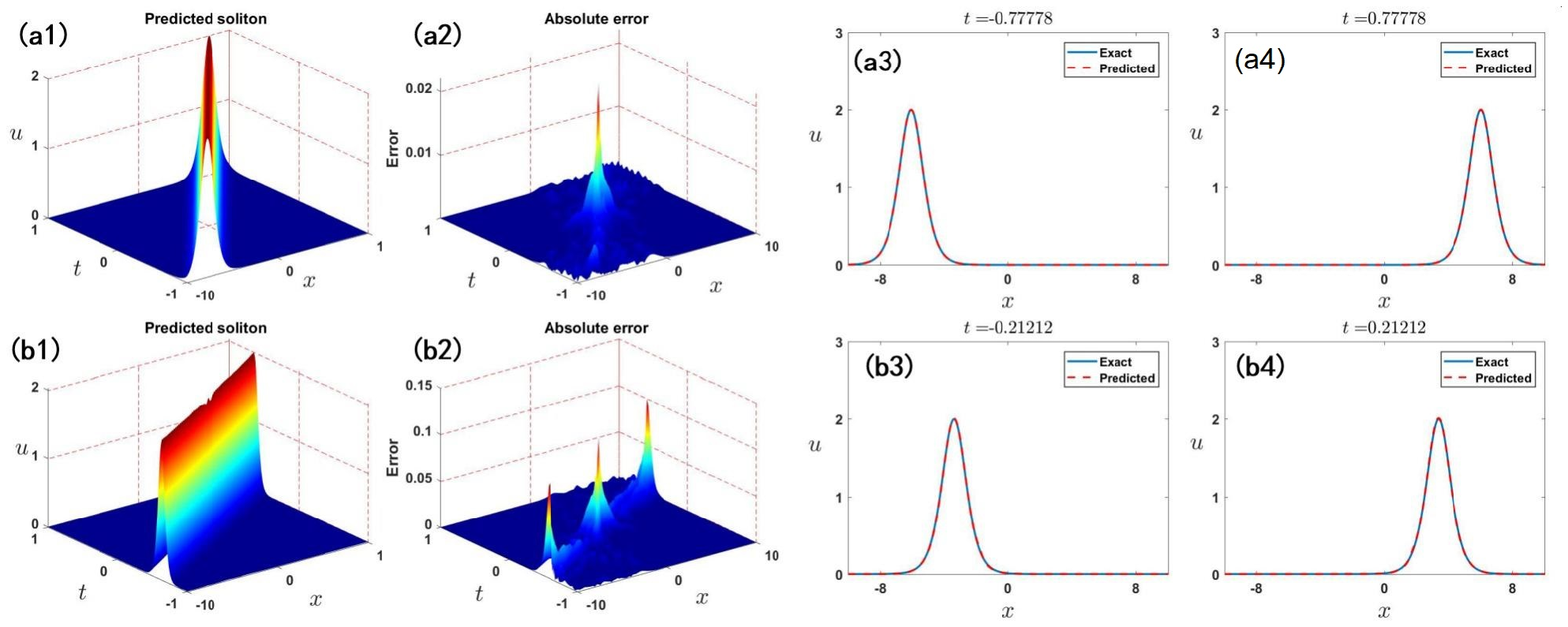}}}\hspace{-0.35in}
\vspace{0.1in}
\caption{(a1)-(a4) best and (b1)-(b4) worst performances of FNO with the Sigmoid$(x)$ activation function.}
  \label{sigmoidM}
\end{figure*}

\begin{figure*}[!t]
    \centering
\vspace{-0.1in}
 {\scalebox{0.75}[0.75]{\includegraphics{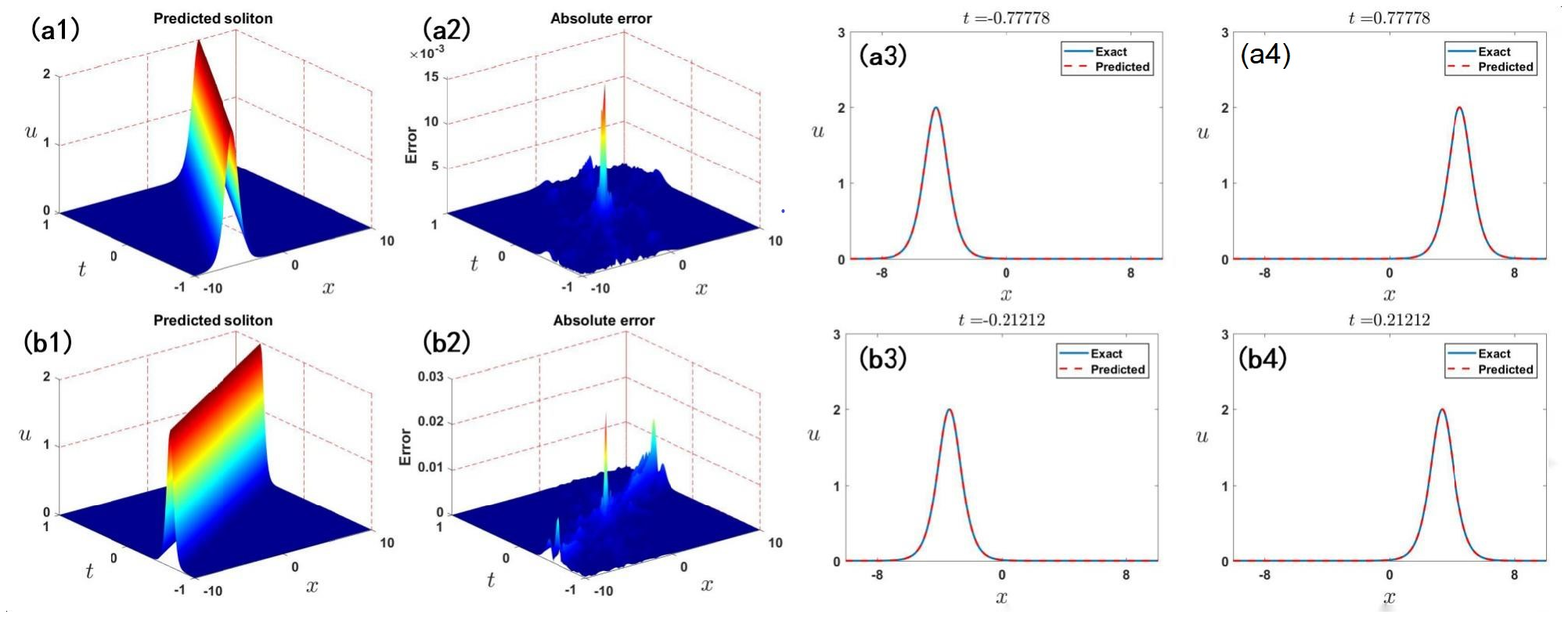}}}\hspace{-0.35in}
\vspace{0.1in}
\caption{(a1)-(a4) best and (b1)-(b4) worst performances of FNO with the Swish$(x)$ activation function. }
  \label{swishM}
\end{figure*}

\begin{figure*}[!t]
    \centering
\vspace{-0.1in}
 {\scalebox{0.75}[0.75]{\includegraphics{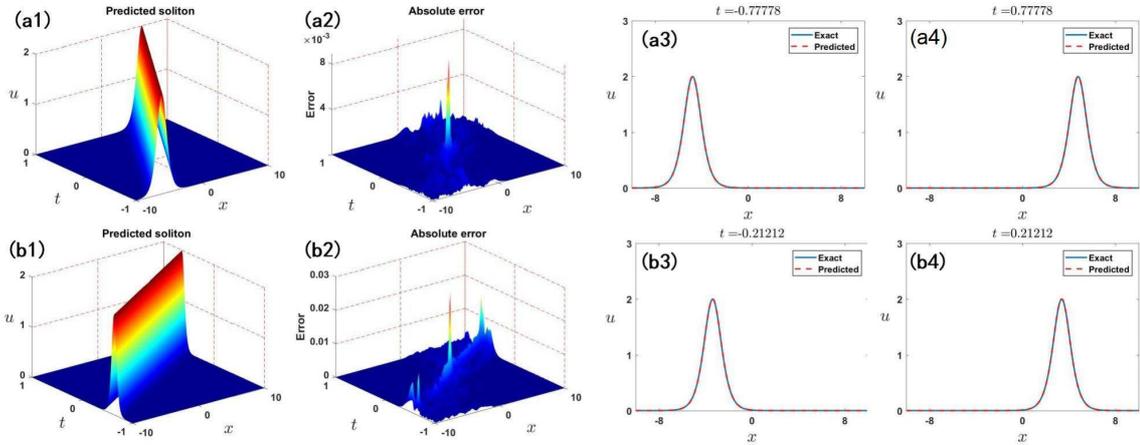}}}\hspace{-0.35in}
\vspace{0.1in}
\caption{(a1)-(a4) best and (b1)-(b4) worst performances of FNO with the $x\tanh(x)$ activation function.}
  \label{xtanhM}
\end{figure*}

It is well known that the nonlinear activation function is an essential part of a neural network, and can  improve its expressive power substantially. Therefore, different types of activation functions have a strong impact on the representational power of the network. In this section, we mainly consider the influence of activation function  on the performance of FNO scheme. We consider the following nonlinear activation functions: Relu$(x)$, Sigmoid$(x)$, Swish$(x)$ and $x\tanh(x)$, respectively. We choose the case of fKdV in subsection 3.1 as the baseline. It's noted that the function $x\tanh(x)$ is firstly introduced into the neural network as the nonlinear activation function.

We display the performances of different types of nonlinear activation functions in Figs.~\ref{reluM}-\ref{xtanhM}. The best and worst performances of FNO with the Relu$(x)$ activation function are exhibited in the first row and second row of  Fig.~\ref{reluM}. And the latter figures also show the influences of  activation functions on the FNO scheme. It can be concluded that $x\tanh(x)$ and Swish$(x)$ activation functions perform well that Relu$(x)$ and Sigmoid$(x)$ functions. The conclusion can also be obtained from the  worst, mean and best relative $L_2$ errors listed in Table.~\ref{Com1}. The corresponding errors of $x\tanh(x)$ and Swish$(x)$ functions are less than those of Relu$(x)$ and Sigmoid$(x)$ functions. And the worst  performance of FNO mainly occurs in the big $\epsilon$ while the best  performance of FNO occurs when $\epsilon$ is small.

\begin{table}
\centering
\caption{Comparisons between the depths of $P$ and relative $L_2$ error.}
\vspace{0.05in}
\begin{tabular}{c|cccc}
\hline\hline
\rule{0pt}{13pt}\diagbox{Depth}{Error}               & \qquad Worst                     &\qquad Mean                        & \qquad Best
 \\ \hline
\rule{0pt}{13pt} 8                     &  \qquad 8.514e-3                 &  \qquad 2.711e-3     & \qquad  1.745e-3                                     \\ \\
12                          & \qquad 5.608e-3    &  \qquad2.128e-3        & \qquad 1.526e-3             \\ \\
16                        & \qquad 5.605e-3             & \qquad 1.595e-3                   &  \qquad 1.062e-3             \\ \\
20                         & \qquad 5.392e-3              &  \qquad 1.421e-3                & \qquad 1.024e-3             \\[1ex]
\hline\hline
\end{tabular}
\label{Com2}
\end{table}
\begin{figure*}[!t]
    \centering
\vspace{-0.1in}
 {\scalebox{0.75}[0.75]{\includegraphics{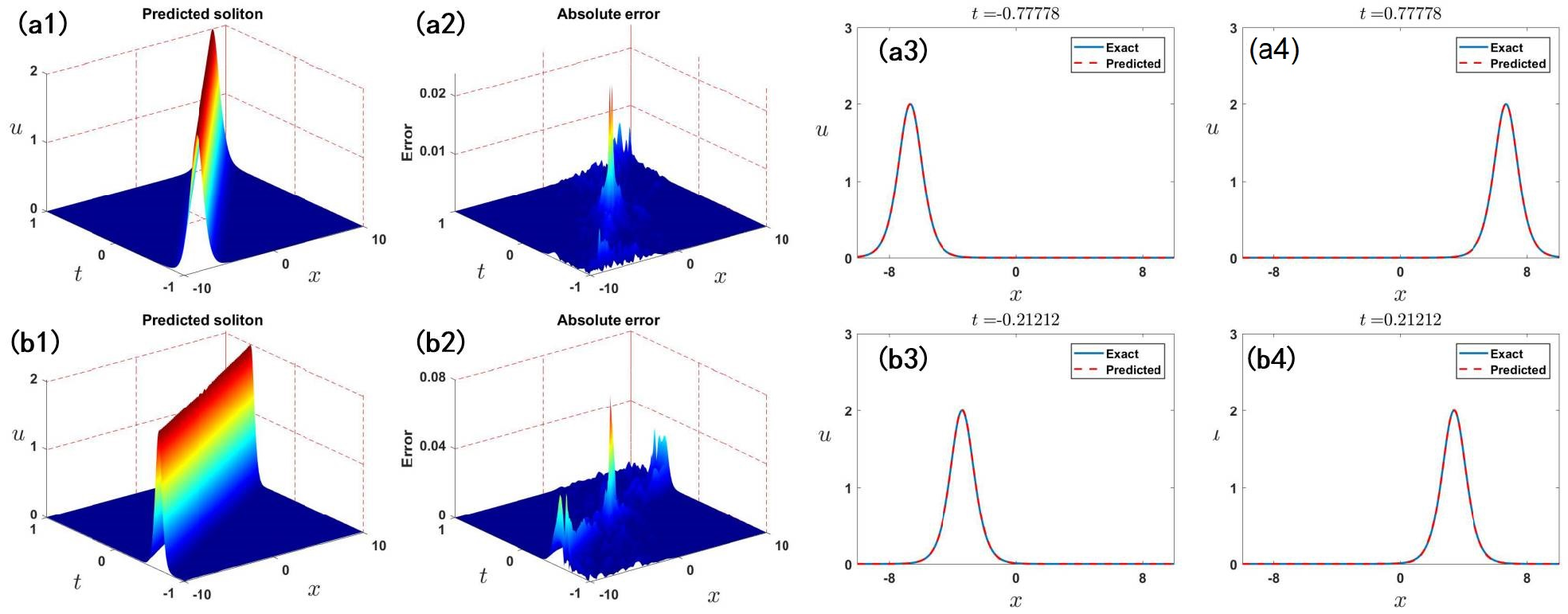}}}\hspace{-0.35in}
\vspace{0.1in}
\caption{(a1)-(a4) best and (b1)-(b4) worst performance of FNO with 8-layer fully connected $P$.}
  \label{depth8M}
\end{figure*}

\begin{figure*}[!t]
    \centering
\vspace{-0.1in}
 {\scalebox{0.75}[0.75]{\includegraphics{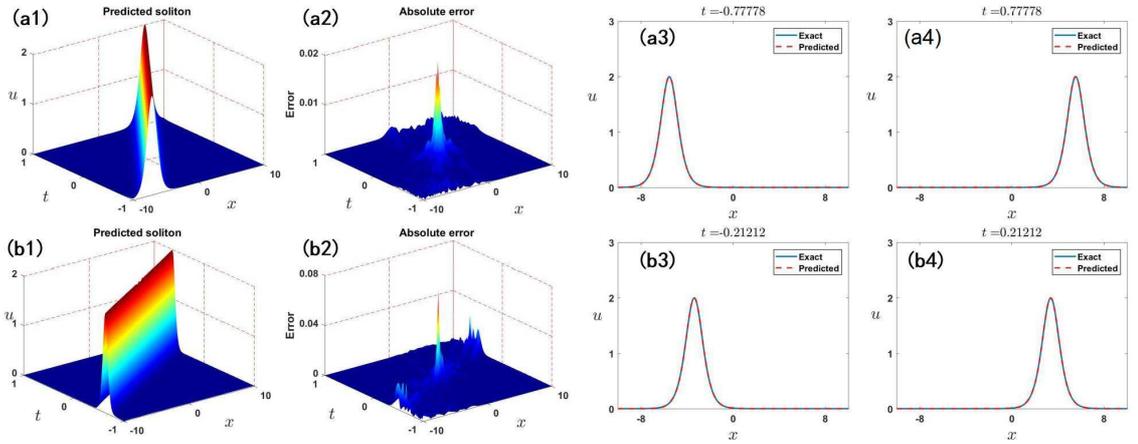}}}\hspace{-0.35in}
\vspace{0.1in}
\caption{(a1)-(a4) best and (b1)-(b4) performances of FNO with a 12-layer fully connected $P$. }
  \label{depth12M}
\end{figure*}

\begin{figure*}[!t]
    \centering
\vspace{-0.1in}
 {\scalebox{0.75}[0.75]{\includegraphics{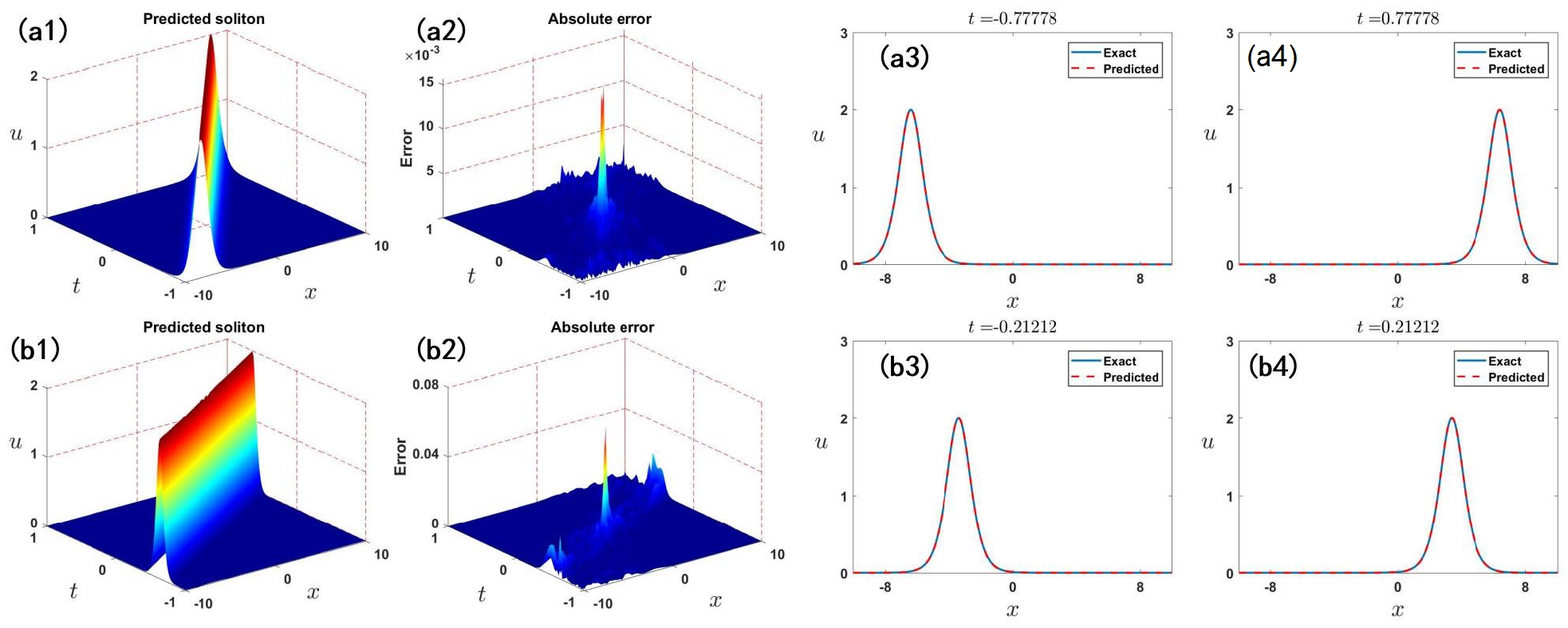}}}\hspace{-0.35in}
\vspace{0.1in}
\caption{(a1)-(a4) best and (b1)-(b4) worst performances of FNO with a 16-layer fully connected $P$. }
  \label{depth16M}
\end{figure*}

\begin{figure*}[!t]
    \centering
\vspace{-0.1in}
 {\scalebox{0.75}[0.75]{\includegraphics{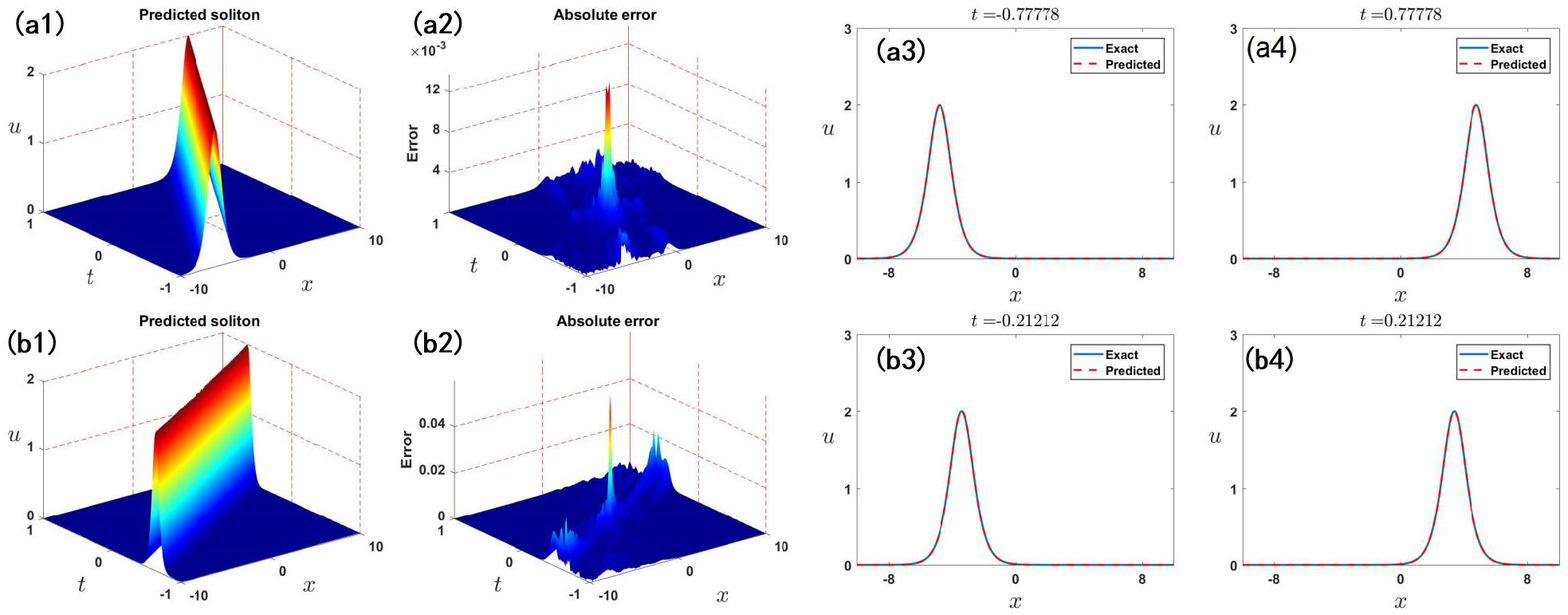}}}\hspace{-0.35in}
\vspace{0.1in}
\caption{(a1)-(a4) best and (b1)-(b4) worst performances of FNO with a 20-layer fully connected $P$. }
  \label{depth20M}
\end{figure*}

\subsection{The influence of  depth for the fully connected layer $P$ }

In order to investigate the effect of the depth of the fully connected layer $P$ on the performance of FNO, different depths for $P$  are studied in this subsection. The depths of the fully connected layer $P$ are selected as 8, 12, 16 and 20, respectively. We also choose the case of fKdV in subsection 3.1 as the baseline.

We display the performances of different depths for the fully connected layer $P$ in Figs.~\ref{depth8M}-\ref{depth20M}. The best and worst performances of FNO with 8-layer fully connected layer $P$ are exhibited in the first and second rows of  Fig.~\ref{depth8M}. And the latter figures also show the influences of  depths on the FNO scheme. It can be concluded that with the increasing depth of layer $P$, the training effect tends to be improved.  The conclusion can also be obtained from the  worst, mean and best relative $L_2$ errors listed in Table.~\ref{Com2}. The corresponding errors of 20-layer are  less than that of 8-, 12- and 16-layer. And the worst  performance of FNO mainly occurs in the big $\epsilon$ while the best  performance of FNO occurs when $\epsilon$ is small.

\section{Conclusions and discussions}

 In conclusion, we have effectively studied the data-driven soliton mappings of the fKdV, fmKdV and fsineG equations by means of the Fourier neural operator method.  Moreover, the influences of critical factors (e.g., nonlinear activation functions and depths of the fully connected layer $P$) have been analyzed to test the learning ability of the deep neural networks.  The results proved that the FNO algorithm is powerful in searching for the fractional soliton mapping between two infinite-dimensional spaces. The results obtained in this paper are useful to further analyze the data mapping and neural operator networks solving other fractional PDEs.

\v\v\noindent{\bf Acknowledgments} \v

The work  was supported by the NSFC under Grant No. 11925108.

\end{document}